\documentclass[conference]{IEEEtran}
\IEEEoverridecommandlockouts
\usepackage{mathrsfs}
\usepackage{amsfonts}
\usepackage{ifpdf}
\usepackage{cite}
\usepackage{array}
\usepackage{booktabs}
\usepackage{setspace}

\ifCLASSINFOpdf
\usepackage[pdftex]{graphicx}
\else \fi
\usepackage[cmex10]{amsmath}
\usepackage{array}
\usepackage{mdwmath}
\usepackage{mdwtab}
\usepackage{eqparbox}
\usepackage[tight,footnotesize]{subfigure}
\usepackage{algorithmic}
\usepackage{algorithm}
\usepackage{amsmath}
\usepackage{epsfig}
\usepackage{ae}
\usepackage{amssymb}
\usepackage{times}
\usepackage{amsmath}   % add

\usepackage{xcolor}

\usepackage{amsthm} % Let Lemma, Theorem, and Proposition be italic.

\usepackage{supertabular} % add

\usepackage{multirow}%用于表格
\newlength\savewidth
\newcommand\shline{\noalign{\global\savewidth\arrayrulewidth
                            \global\arrayrulewidth 1.5pt}%
                   \hline
                   \noalign{\global\arrayrulewidth\savewidth}}

\begin{document}
%----------------------------------------Make Title -----------------------------------------

%-------------------------------------------------------------------------------------------
%\title{Joint Mode Selection, Admission Control, Partner Assignment, and Power Allocation in Underlay SCMA Device-to-Device Networks}
\title{Adaptive F-FFT Demodulation for ICI Mitigation in Differential Underwater Acoustic OFDM Systems}
\author{
     \IEEEauthorblockN{
        Jihui Qiu\textsuperscript{1,2}, Yuzhou Li\textsuperscript{1,2}, Yunlong Huang\textsuperscript{1}, Yimeng Wang\textsuperscript{1}, and Lingyu Gu\textsuperscript{1}\\
    }
    \IEEEauthorblockA{
        \textsuperscript{1}Huazhong University of Science and Technology, Wuhan, 430074, P. R. China\\
        \textsuperscript{2}State Key Laboratory of Integrated Services Networks (Xidian University), Xi'an, 710071, P. R. China\\
        \{jihuiqiu, yuzhouli, ylhuang, wangyim, lingyugu\}@hust.edu.cn}

\thanks{
This work was supported in part by the National Science Foundation of China under Grant 61971462, Grant 61831013, and Grant 61631015 and the State Key Laboratory of Integrated Services Networks (Xidian University) under Grant ISN19-09.
}
}
\maketitle
\IEEEpeerreviewmaketitle
%---------------------------------------Make Abstract--------------------------------------
\begin{abstract}
This paper addresses the problem of frequency-domain inter-carrier interference (ICI) mitigation for differential orthogonal frequency-division multiplexing (OFDM) systems. The classical fractional fast Fourier transform (F-FFT), adopting the fixed sampling interval, would suffer from the limited accuracy of ICI mitigation and low adaptability in dynamic Doppler spread. To target the above challenges, we propose an adaptive fractional Fourier transform (A-FFT) demodulation method, in which an estimation algorithm based on the coordinate descent approach is designed to compute the fiducial frequency offset without increasing pilots. By means of compensating ICI at fractions of the fiducial frequency offset adapted to the time-varying Doppler shift, the A-FFT has the capability of tracking Doppler fluctuations over the underwater acoustic channels, thus extending the application range of frequency-domain ICI mitigation. Simulation results show that the A-FFT is significantly superior to the existing classical methods, the partial fast Fourier transform (P-FFT) and the F-FFT, for both medium and high Doppler factors and large carrier numbers in terms of the mean squared error (MSE). Numerically, the MSE of the A-FFT is reduced by $\bf{39.88\%-72.14\%}$ compared to that of the F-FFT with the input signal-to-noise ratio ranging from 10 dB to 30 dB at a Doppler factor of $\bf{2.5\times 10^{-4}}$ and a carrier number of 1024, while the P-FFT even cannot work well.

\end{abstract}
%--------------------------------------------- Make Key words-----------------------------

\section{Introduction}
A key issue for the coherent detection to accurately recover the transmitted data is precise channel estimation. Nevertheless, pilot-aided acquisition of the channel state information over the  hostile underwater acoustic (UWA) channel tends to be resource-exhaustive and high-complexity \cite{ChannelEstimateThree,ChannelEstimateFour,channelEstimation,DopplerRateRange}. In view of this, the differentially coherent detection has become a promising alternative scheme, since it has the potential to eliminate the need for channel estimation. Contrasted to the coherent counterpart, it relies steadily on the strong coherence between adjacent carriers or blocks to achieve the above goals. With this in mind, considering an orthogonal frequency-division multiplexing (OFDM) system with differential encoding in frequency, the largest possible number of carriers should be used to approximate the channel invariance across neighboring carriers, i.e., ensure the strong coherence among carriers. However, the larger carrier number, equivalent to the narrower carrier spacing, would in turn cause the system to be more sensitive to the frequency offset, and the resulting inter-carrier interference (ICI) dominates more severe detection errors. To this end, it is important to use as many carriers as possible to support the strong coherence between adjacent carriers, and yet maintain the ICI to a low level in differential OFDM systems.

Lots of thorough researches on ICI mitigation have been conducted from different perspectives. Among them, a class of approaches with multiple fast Fourier transform (FFT) demodulation outputs have emerged as a promising solution and tested for differential OFDM systems \cite{fullProposed}. In general, these methods focus on the combination of pre-FFT filtering processing and post-FFT weight compensation. According to the processing domain for ICI mitigation, existing methods can be divided into two categories:
\begin{itemize}
\item Time-domain ICI mitigation methods divide the received signal in the time domain to reduce the channel variations, thus converting the fast time-varying channel into several quasi-static subchannels, represented by the partial fast Fourier transform (P-FFT) \cite{PFFTOne,PFFTTwo,PFFTThree,PFFTFour}.

\item Frequency-domain ICI mitigation methods utilize spectrum oversampling to compensate for Doppler distortions in the frequency domain \cite{frequencyResample}, thus reducing ICI and the complexity of full channel estimation, with the fractional fast Fourier transform (F-FFT) and its revised versions being the representative ones \cite{firstProposed,DFFFT}.
\end{itemize}

A large amount of theoretical and analytical results demonstrate that frequency-domain ICI mitigation methods usually outperform the time-domain one. Based on this fact, we in this paper focus on the frequency domain to alleviate ICI. As far as we know, the F-FFT was first introduced in \cite{firstProposed}, compensating frequency offset at the half of the carrier spacing. The follow-up work in \cite{fullProposed} further extended it to a multichannel framework, coupled with a gradient scaling method to improve the original weight compensation algorithm. Furthermore, a decision fractional FFT was designed in \cite{DFFFT} to enhance the performance of the F-FFT through symbol rebuilding. Although these works have confirmed the effectiveness of the F-FFT in compensating ICI for differential OFDM, the performance would degrade when the dynamic range of Doppler deviation is large. This is because the classical F-FFT, adopting fractional multiples of the carrier spacing for frequency oversampling, has the limited range of ICI compensation when the size of the combiner is fixed. As a result, the performance ceiling can be observed due to the poor adaptability  in dynamic Doppler spread.

To this end, in this paper we propose a different approach that targets time-varying ICI for acoustic channels with severe Doppler fluctuations. Unlike the fixed carrier spacing of the existing F-FFT, the compensation range of the proposed method at each frame would fluctuate with Doppler. Namely, it is supposed to reset the fiducial frequency offset adapted to the current Doppler spread before demodulating each data frame, by fractional multiples of which to shift spectrum for compensating ICI. In this manner, an adaptive fiducial frequency offset estimation algorithm is designed employing the coordinate descent approach with no increase in additional pilot overhead. Combining with the designed algorithm, the proposed method has the capability of tracking dynamic Doppler spread over the UWA channel, thus extending the application range of frequency-domain ICI mitigation, which is referred to as adaptive fractional Fourier transform (A-FFT). Simulation results demonstrate that the A-FFT significantly outperforms the P-FFT and the F-FFT for both medium and high Doppler factors and large carrier numbers in terms of the data detection mean square error (MSE).

The remainder of this paper is organized as follows. In Section~\ref{Section:ModelAndDemodulation}, we present the system model for the UWA differential OFDM system and the principle of F-FFT demodulation. Section~\ref{Section:Estimate} describes the performance influencing factor of the existing F-FFT and the details of the proposed A-FFT demodulation. Simulation results are presented in Section ~\ref{Section:SimulationResults} to evaluate the performance of the proposed method. Finally, conclusions are summarized in Section~\ref{Section:Conclusions}.

\section{System Model And F-FFT Demodulation} \label{Section:ModelAndDemodulation}

\subsection{System Model} \label{Subsection:SystemModel }
Let us consider a differential OFDM system with $K$ carriers, in which the transmitted signal is modeled as
\begin{equation} \label{Eq:transmittedSignal}
s(t)={\rm{Re}}\left \{\sum_{k=0}^{K-1}d_ke^{j2\pi f_k t}\right\},t\in[0,T]
\end{equation}
where $\Delta f$ is the carrier spacing and $T=1/\Delta f$ is the block duration. The data symbol $d_k$, modulated on the $k$-th carrier of frequency $f_k=f_0+k\Delta f$ with the lowest carrier $f_0$, is differentially encoded in the frequency domain, following as
\begin{equation} \label{Eq:diffEncoding}
d_k=
\begin{cases}
b_kd_{k-1}, & 1 \leq k \leq K-1 \\
c, & k = 0
\end{cases}
\end{equation}
where $c$ is a known symbol for both the transmitter and the receiver, and $b_k$ is the original data symbol drawn from the Q-ary unit-amplitude phase-shift keying (PSK) constellation alphabet set.

Assuming that a multipath acoustic channel with the path gain $h_l(t)$ and delay $\tau _l(t) $ corresponding to each path is modeled as
\begin{equation} \label{Eq:recievedSignal}
h(t)=\sum_{l=0}^{L-1}h_l(t)\delta \left(t-\tau _l(t) \right)
\end{equation}
the received signal in passband is then expressed as
\begin{equation} \label{Eq:recievedSignal}
r(t)=\sum_{l=0}^{L-1}h_l(t)s\left(t-\tau _l(t) \right)+n(t)
\end{equation}
where $n(t)$ is the additive complex noise.

After frame synchronizing, coarse resampling, down-shifting and guard interval discarding, the equivalent received signal in baseband can be written as
\begin{equation} \label{Eq:recievedBaseSignal}
v(t)=\sum_{k=0}^{K-1}H_k(t)d_k e^{j2\pi k \Delta ft}+z_k(t)
\end{equation}
in which $ H_k(t)=\sum_{l=0}^{L-1}h_l(t)e^{-j2\pi f_k \tau_l(t)}$ and $ z_k(t)$ denote the channel coefficient and the equivalent noise, respectively.

\subsection{F-FFT Demodulation} \label{Subsection:FFFTDemodulation}

In classical fractional FFT demodulation, the received OFDM signal is shifted $I$ times by fractions of the carrier spacing and a Fourier transform is performed on each frequency shifted signal. The output of the Fourier transform for the $k$-th subcarrier and the $i$-th frequency shifted signal, henceforth called the F-FFT outputs, can be expressed as
\begin{equation} \label{Eq:convFFT}
y_{k,i}=\int_{0}^{T}v(t)e^{-j2 \pi (k+\frac{i}{I})\Delta f t},i=0, 1,\ldots,I-1.
\end{equation}

The F-FFT outputs are then combined to form the final demodulated signal
\begin{equation} \label{Eq:combiner}
x_k={\bf{w}}_{k}^{H}{\bf{y}}_k
\end{equation}
where ${\bf{w}}_{k}$ is the vector of combiner weights on the $k$-th carrier and its total length is denoted by $M$. The demodulator output vector ${\bf{y}}_k $ consists of the original FFT output $y_{k,0}$ as the central element and $(M-1)$ adjacent elements shifted by multiples of positive and negative $\Delta f/I$, represented as
\begin{equation} \label{Eq:demodulateOut}
\begin{aligned}
{\bf y}_{k}=\left[\ldots, y_{k-1,1},\ldots,y_{k-1,I-1},y_{k,0},\qquad \qquad\right. \hfill\cr\hfill\left.y_{k,1},\ldots ,y_{k,I-1},\ldots\right]^{T}.
\end{aligned}
\end{equation}
Namely, the length of ${\bf y}_k$ would vary with that of the combiner weight vector ${\bf w}_k$.

\section{A-FFT Demodulation} \label{Section:Estimate}
In this section, we introduce the concept of fiducial frequency offset and point out its impact on performance of the existing F-FFT. Then, the A-FFT demodulation, combined with the fiducial frequency offset estimation algorithm, is proposed to deal with the disadvantages of the existing method.

\subsection{Fiducial Frequency Offset} \label{Subsection:analysis}

In the existing F-FFT demodulation, the dimension $M$ of the combiner vector is generally selected as $M=2I-1$. In this case, Eq. (\ref{Eq:convFFT}) can be rewritten as
\begin{equation} \label{Eq:FFFT}
z_{k,a}=\int_{0}^{T}v(t)e^{-j2 \pi (k\Delta f+\frac{a}{A+1} f_e) t}dt,a=0, \pm 1,\ldots,\pm {A}
\end{equation}
in which $A=I-1$, and $f_e$ is referred to as the fiducial frequency offset in this paper and equals to the carrier spacing numerically, thus forming the corresponding demodulator output vector ${\bf{z}}_k $ as
\begin{equation} \label{Eq:demodulateOutput}
{\bf z}_{k}=\left[z_{k,-A},\ldots ,z_{k,-1},z_{k,0},z_{k,1},\ldots ,z_{k,A}\right]^{T}
\end{equation}
and the demodulated signal $x_k$ remains as
\begin{equation} \label{Eq:combinerAct}
x_k={\bf{w}}_{k}^{H}{\bf{z}}_k.
\end{equation}

By means of spectral sampling at fractions of the fixed fiducial frequency offset, the existing F-FFT makes compensation of any Doppler shift in the range $\left [-A\Delta f/(A +1),A\Delta f/(A+1) \right ]$ feasible. However,  due to the time-varying characteristics of the underwater acoustic channel, the Doppler spread has a large dynamic range \cite{ChannelEstimateFour}. In this case, taking the carrier spacing as a fixed frequency offset greatly degrades the dynamic Doppler adaptability of the method, which in turn leads to poor interference mitigation performance. With this in mind, the largest possible fiducial frequency offset seems to be applied to compensate for the Doppler shifts when $A$ is fixed.

\begin{figure}[t]
\centering \leavevmode \epsfxsize=3.5 in  \epsfbox{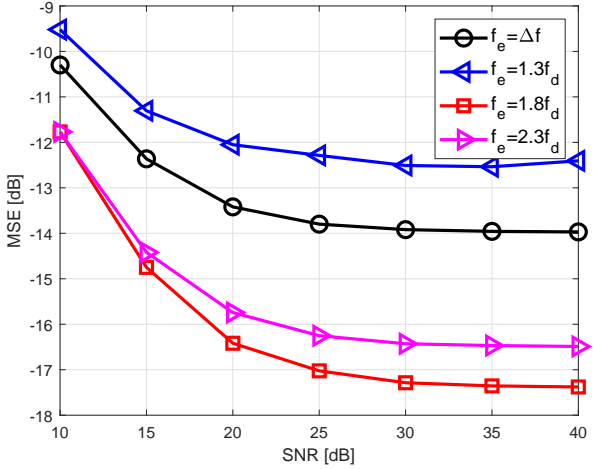}
\centering \caption{The MSE performance of F-FFT demodulation with different fiducial frequency offsets $f_e$, where the Doppler shift at the center frequency $f_d=$3.25 $\rm{Hz}$, $\Delta f=$4.88 $ \rm{Hz}$.} \label{Fig:feInfluence}
\end{figure}

Fig.~\ref{Fig:feInfluence} shows the MSE performance of F-FFT demodulation with different fiducial frequency offsets $f_e$. It can be intuitively seen that the data detection MSE is not optimal when the fiducial frequency offset is set to the carrier spacing. And as expected, the system performance improves as $f_e$ increases. However, the MSE rises when $f_e$ is larger than a certain threshold, that is, unreasonable frequency oversampling causes the system performance to be impaired instead. Therefore, from the practical point of view, a suitable value of $f_e$ is supposed to be determined in advance. Based on this observation, we take the estimation of $f_e$ as the breakthrough point, and deeply analyze the analytical quantitative relationship between MSE and $f_e$. The corresponding estimation algorithm is then proposed to maximize the ICI mitigation performance of the system in this paper.

\subsection{Adaptive Estimation Algorithm} \label{Subsection:estimation}
Considering the narrowband characteristics of the received signal after resampling, it can be assumed that the fiducial frequency offset $f_e$ within a frame is approximately equal. Namely, the optimal $f_e$ of the current frame needs to be determined before demodulating the signal frame by frame according to the preamble.

Using the first $P$ carriers of the first block as pilot symbols, the composite MSE is formed following as
\begin{equation} \label{Eq:mse}
\begin{aligned}
E({\bf{w}}_{k_0},\ldots,{\bf{w}}_{k_{P-1}},f_e)
&= \sum_{k\in  {\mathcal{K}}_{\rm {P}}} \left | e_k\right | ^{2}\\
&= \sum_{k\in  {\mathcal{K}}_{\rm {P}}} \left | b_k - \hat{b}_k \right |^{2}\\
\end{aligned}
\end{equation}
where ${\mathcal{K}}_{\rm {P}}=\{k_0,k_1,\ldots,k_{P-1}\}$ is the set of pilot carriers, and $\hat{b}_k$ is the estimated symbol after differentially coherent detection, calculated by
\begin{equation} \label{Eq:DiffDetection}
\hat{b}_k = \frac{x_k}{x_{k-1}} = \frac{{\bf w}_{k}^{H}{\bf z}_{k}}{{\bf w}_{k-1}^{H}{\bf z}_{k-1}}.
\end{equation}

To derive the optimal fiducial frequency offset $\hat{f}_e$  in light of minimizing MSE in data detection, i.e.,
\begin{equation} \label{Eq:mmse}
\hat{f}_e=\arg \underset{f_e}{\min} \left \{{E\left ({\bf{w}}_{k_0},\ldots,{\bf{w}}_{k_{P-1}},f_e\right)}\right \},
\end{equation}
we take a coordinate descent approach where the composite MSE of data detection (\ref{Eq:mse}) is used to guide the estimation of $f_e$. With $f_e$ independent of ${\bf w}_k$, the partial derivative of the MSE with respect to ${f}_{e}$ is given by
\begin{equation} \label{Eq:feGrad}
\begin{aligned}
{\partial E\over \partial {f}_{e}}\!=-2\pi \sum_{k\in  {\mathcal{K}_p} }\!{\rm Im}\!\left\{{{\bf w}_k^H \left ({\pmb \beta} \circ \tilde{\bf z}_k \right )x_{k-1}\over (x_{k-1})^{2}}e_k^{\ast}\qquad \qquad
\right.\hfill\cr\hfill\left.\vphantom{{\rm Im}}
 -{{\bf w}_{k-1}^H \!\left ({\pmb \beta} \circ \tilde{\bf z}_{k-1} \right )x_k\over (x_{k-1})^{2} }
e_k^{\ast}\right\}
\end{aligned}
\end{equation}
in which
\begin{subequations}\label{Eq:subEqu}
\begin{align}
& \pmb \beta=\left [-A,\ldots,-1,0,1,\ldots,A \right ]^{T}\\
& \tilde{z}_{k,a}=\int_{0}^{T}tv(t)e^{-j2 \pi (k\Delta f+\frac{a}{A+1} f_e) t}dt\\
& \tilde{\bf z}_{k}=\left[\tilde{z}_{k,-A},\ldots ,\tilde{z}_{k,-1},\tilde{z}_{k,0},\tilde{z}_{k,1},\ldots ,\tilde{z}_{k,A}\right]^{T}
\end{align}
\end{subequations}
and $``\circ"$ denotes the Hadamard product.

Furthermore, we define
\begin{equation} \label{Eq:demodulateSymbol}
\begin{aligned}
\tilde{x}_k & ={\bf w}_k^H \left ({\pmb \beta} \circ \tilde{\bf z}_k \right )
\end{aligned}
\end{equation}
and the gradient $\gamma$ of the fiducial frequency offset is then written as
\begin{equation} \label{Eq:feGradre}
\gamma  =\sum_{k\in  {\mathcal{K}_p} } {\rm Im}\left \{
{\tilde{x}_kx_{k-1}-\tilde{x}_{k-1}x_k\over (x_{k-1})^{2} }e_k^{\ast}
\right \}
\end{equation}
thus $f_e$ can be calculated iteratively following as
\begin{equation} \label{Eq:feRecurion}
f_e^{iter}(j+1)=f_e^{iter}(j)+\mu_ {f_e} \gamma^{iter}(j)
\end{equation}
in which $f_e^{iter}(j)$ represents the $j$-th iterative value of the current $f_e^{iter}$ inner loop. Among them, $f_e^{iter}(0)=f_e^{iter}$ and $f_e^{iter}$ indicates the fiducial frequency offset of the $iter$-th outer cycle with the initial value $f_e^0$, generally set to $\Delta f$ or several times the Doppler shift at the center frequency.

\begin{figure}[t]
\centering \leavevmode \epsfxsize=3.5 in  \epsfbox{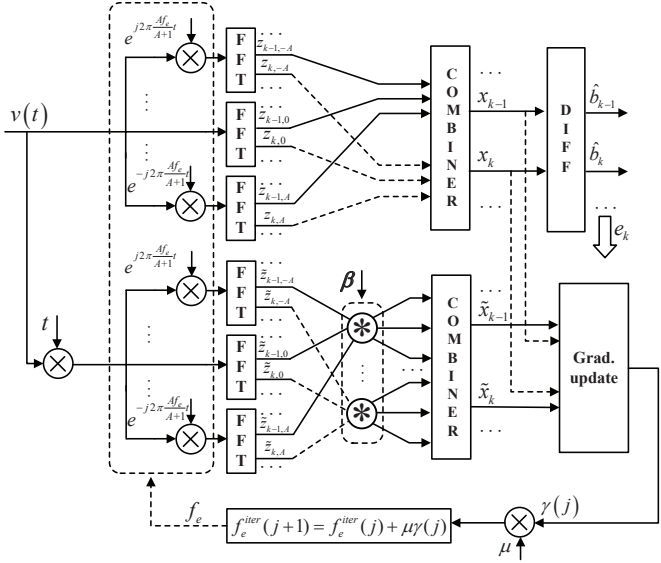}
\centering \caption{Block diagram of the fiducial frequency offset estimation algorithm, where $``*"$ means the Hadamard product operation in this figure, $``\rm{DIFF}"$ and $``\rm{Grad. update}"$ denote the differentially coherent detection and gradient update, respectively.} \label{Fig:blockDigram}
\end{figure}

Taking into account the dependence $E$ on ${\bf{w}}_k$, we also consider the partial derivative of the MSE with respect to ${\bf w}_{k_p}$, given by
\begin{equation} \label{Eq:wkGrad}
{\partial E\over \partial {\bf w}_{k_p}^{\ast}} =-\underset{{\bf g}_{k_p}}{\underbrace{ \left({{\bf z}_{k_p}\over x_{k_{p-1}} }e_{k_p}^{\ast}-{{\bf z}_{k_p}x_{k_{p+1}}\over (x_{k_p})^{2} }e_{k_{p+1}}^{\ast}\right)}}
\end{equation}
and the weights can be obtained by
\begin{equation} \label{Eq:wkRecurion}
{\bf w}_{k_p}^{iter}(j+1)={\bf w}_{k_p}^{iter}(j)+\mu _{k_p} {\bf g}_{k_p}^{iter}(j)
\end{equation}
with a maximum number of iterations $N_I$ or a predefined error threshold $\eta$, and ${\bf w}_{k_p}^{iter}(0)={\bf w}_{k_p}^{iter}$ where ${\bf w}_{k_p}^{iter}$ denotes the combiner weight ${\bf w}_{k_p}$ of the $iter$-th outer loop. Similarly, $j$ represents the number of iterations in the inner loop of the current weight vector ${\bf w}_{k_p}^{iter}$. For the iteration $iter=0$, ${\bf w}_{k_p}^{0}$ is set to the $(2A+1)\times 1$ vector ${[\ldots,0,1,0,\ldots]}^{T}$. Fig.~\ref{Fig:blockDigram} shows the block diagram of the estimation algorithm, in which $\tilde{x}_k$ is obtained using the F-FFT with the same weight as $x_{k}$ and a set of Hadamard product multipliers, applied to the samples of $tv(t)$.

\begin{algorithm}[!t]
\caption{Fiducial Frequency Offset Estimation Algorithm}
\begin{algorithmic}[1] \label{Algorithm:AdaptiveAlgorithm}
\STATE $\bf{Input:}$
$v(t)$, pilot symbols $b_k$, $k \in {\mathcal{K}_P}$
\STATE $\bf{Output:}$ $\hat{f_e}$
\STATE $\bf{Initialization}$
\begin{itemize}
  \item Set parameters ${\bf w}_{k_0}^{0},\ldots,{\bf w}_{k_{P-1}}^{0},f_e^0$
  \item Set $\eta$, $N_I$, step size $\mu_{{\bf w}_{k_0}},\ldots,\mu_{{\bf w}_{k_{P-1}}},\mu_{f_e}$
\end{itemize}
\WHILE {$(iter \leqslant N_I$)}
\STATE Given ${\bf w}_k^{iter}=\left ({\bf w}_{k_0}^{iter},\ldots,{\bf w}_{k_{P-1}}^{iter} \right )$, choose an index $p \in \{0,\ldots,P-1\}$ and compute a new iterate
\STATE $ \quad {\bf w}_k^{iter+1}=\left ({\bf w}_{k_0}^{iter+1},\ldots,{\bf w}_{k_{P-1}}^{iter+1} \right )$
\STATE satisfying
\STATE  $\quad {\bf w}_{k_p}^{iter+1} \in \arg \underset{{\bf w}_{k_p}}{\min} \left \{E \left({\bf w}_{k_0}^{iter},\ldots,{\bf w}_{k_p},\ldots, f_e^{iter}\right)\right \}$
\STATE  $\quad {\bf w}_{k_i}^{iter+1}=w_{k_i}^{iter}, \quad  \forall i\neq p$
\STATE $f_e^{iter+1}\in \arg \underset{f_e}{\min} \left \{{E\left ({\bf{w}}_{k_0}^{iter+1},\ldots,{\bf{w}}_{k_{P-1}}^{iter+1}, f_e \right)}\right \}$
\IF {$ 10 \left | \log_{10} {E \left({\bf w}_{k_0}^{iter},\ldots,{\bf w}_{k_{P-1}}^{iter},f_e^{iter}\right) \over E \left({\bf w}_{k_0}^{iter+1},\ldots,{\bf w}_{k_{P-1}}^{iter+1},f_e^{iter+1}\right)}\right |<  \eta $}
\STATE break
\ENDIF
\STATE $ iter=iter+1 $
\ENDWHILE
\RETURN {$\hat{f_e}=f_e^{iter}$}
\end{algorithmic}
\end{algorithm}

It is worth noting that, in order to reduce the additional pilot overhead, the proposed algorithm should reuse exactly the same pilot with the subsequent weight estimation algorithm as much as possible. Therefore, the number of pilots $P$ is not greater than the number required to estimate the weight. Besides, the objective function of formula (\ref{Eq:mse}) is essentially an unconstrained optimization problem. In addition to the coordinate descent method employed in this article, other constraint optimization methods are equally applicable to the fiducial frequency offset estimation. In view of this, we mainly provide a corresponding target model and feasible solution to deal with the poor adaptability of the existing F-FFT under dynamic Doppler. The formal steps of the fiducial frequency offset estimation algorithm are summarized in Algorithm \ref{Algorithm:AdaptiveAlgorithm}.

After estimating the fiducial frequency offset adapted to the current Doppler spread, the subsequent signal processing operations remain as Eq.~(\ref{Eq:FFFT})-(\ref{Eq:combinerAct}) to accurately demodulate the ICI suppressed signal, which is referred to as adaptive fractional Fourier transform in this paper.

\section{Simulation Results And Analysis} \label{Section:SimulationResults}
In this section, we assess the performance of the proposed ICI mitigation method through simulation and compare it to that of other methods, including the P-FFT in \cite{PFFTThree}, the F-FFT and the differentially coherent detection with the conventional FFT (Conv-FFT), in terms of the data detection MSE.

\subsection{Simulation Parameter Settings} \label{Subsection:SimulationParaSet}

\begin{table}[t]
\centering
\caption{\label{Table:SimulationParameters}Summary of the UWA-OFDM parameter settings.}
\begin{supertabular}{l >{}p{3.1cm}}
\shline
Parameters & Values  \\
\hline
Center frequency $f_c$      & 32 kHz\\
\hline
Signal bandwidth $B$        & 12 kHz\\
\hline
Sampling rate $f_s$        & 192 kHz\\
\hline
Sampling interval $T_s$       & 5.208 us\\
\hline
Number of carriers per block $K$   & $2^6 - 2^{11}$\\
\hline
Number of blocks per frame $N$     & $2^2 - 2^7$\\
\hline
Carrier spacing $\Delta f$  & $5.9 - 187.5$ Hz\\
\hline
Block duration $T$          & $5.3 - 170.7$ ms\\
\hline
Guard interval $T_g$        & 16 ms\\
\hline
Modulation type             & QPSK\\
\shline
\end{supertabular}
\end{table}

In the simulation, a differential OFDM system with typical parameter settings is considered, as summarized in Table~\ref{Table:SimulationParameters}. Notably, $K$ and the number of blocks per frame $N$ always satisfy with $KN=2^{13}$. We adopt the statistical channel model provided in \cite{channelModel} to simulate the UWA channel to assist performance comparison and analysis. Moreover, for the fairness of comparison, the number of demodulation segments for the P-FFT, the F-FFT and the A-FFT are all set to 3, i.e., $I=3$ and $A=1$. We employ the same combiner weight estimation algorithm for the above three ICI mitigation methods equipped with the gradient scaling and the thresholding method provided in \cite{fullProposed}, for which the pilot symbols are inserted on the first 200 carriers of the first OFDM symbol in each frame.

\subsection{MSE Versus SNR} \label{Subsection:MSEPerformanceVersusSNR}

\begin{figure}[t]
\centering \leavevmode \epsfxsize=3.5 in  \epsfbox{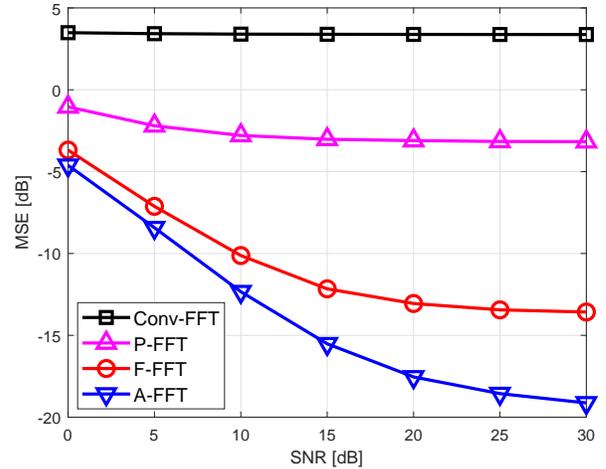}
\centering \caption{Comparisons of the MSE versus input SNRs ranging from 0 dB to 30 dB among different methods, where $K=1024$ and $\alpha= 2.5\times 10^{-4}$.} \label{Fig:MSESNR}
\end{figure}

Fig.~\ref{Fig:MSESNR} illustrates the performance of different methods in terms of data detection MSE with various input signal-to-noise ratio (SNR) at the receiver, in which $K = 1024$ and the Doppler factor $\alpha = 2.5\times 10^{-4}$. The conventional differentially coherent detection method cannot work normally at any SNR, i.e., the performance cannot be improved with the increase of the SNR, implying that increasing the SNR will not have successful compensation for ICI. Coupled with the fact that the MSE performance of the P-FFT is above $-5$ dB, in contrast, the MSE performance of the other two methods show better improvement with SNR. Moreover, the A-FFT reduces the MSE by $39.88\%-72.14\%$ compared to the F-FFT with the SNR ranging from 10 dB to 30 dB.

\subsection{MSE Versus Doppler Factor} \label{Subsection:MSEPerformanceVersusDoppler}
\begin{figure}[t]
\centering \leavevmode \epsfxsize=3.5 in  \epsfbox{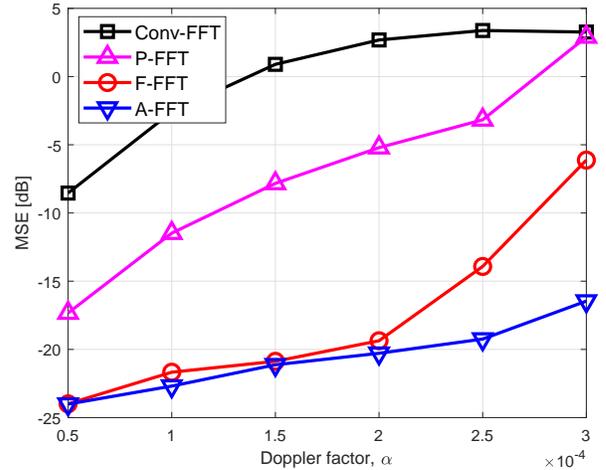}
\centering \caption{Comparisons of the MSE versus the Doppler factor $\alpha$ ranging from $5\times 10^{-5}$ to $3\times 10^{-4}$ among different methods, where $K=1024$ and $\rm{SNR}=30$ dB.} \label{Fig:aMSE}
\end{figure}

Fig.~\ref{Fig:aMSE} demonstrates the MSE performance of the proposed method as a function of the Doppler scaling factor, which ranges between $5\times 10^{-5}$ and $ 3\times 10^{-4}$. Here, $K =1024$ and $\rm{SNR}= 30$ dB. The result asserts that the MSE among all methods increases with the Doppler factor, ultimately limiting detection performance. Nevertheless, the increase of other three methods with multiple FFT demodulation is much slower than that of the conventional receiver, thus expanding the range of tolerable Doppler distortion. Among them, the F-FFT and the A-FFT provide frequency domain oversampling with the preset spacing, which makes it easier to compensate for the serious Doppler effect of the received signal, thus showing stronger anti-Doppler capability. Furthermore, the A-FFT, transforming the frequency shift compensation interval from a fixed spacing to a fractional multiple of the fiducial frequency offset adapted to the dynamic Doppler spread, exhibits better performance, e.g., the MSE can be reduced by $90.74\%$ compared to the F-FFT with $\alpha=3\times 10^{-4}$.

\subsection{MSE Versus the Number of Carriers} \label{Subsection:MSEPerformanceVersusCarrierNum}
\begin{figure}[t]
\centering \leavevmode \epsfxsize=3.5 in  \epsfbox{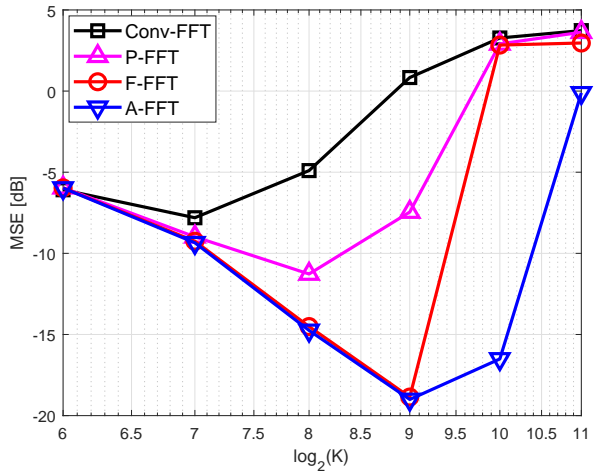}
\centering \caption{Comparisons of the MSE versus the number of carriers $K$ among different methods, where $\rm{SNR}=30$ dB and $\alpha=3\times 10^{-4}$.} \label{Fig:KMSE}
\end{figure}

Fig.~\ref{Fig:KMSE} compares the MSE performance as a function of the number of carriers with $\rm{SNR}=30$ dB and $\alpha = 3 \times10^{-4}$. The results demonstrate the effectiveness of ICI mitigation with the P-FFT, the F-FFT and the proposed A-FFT. These techniques provide an improvement of several dB over conventional method and enable operation with a larger carrier number, effectively increasing the bandwidth efficiency at the same time. Moreover, it can be observed that there exists a deterioration in performance for the OFDM frames when $K$ grows to a certain threshold, which can be explained by shortened carrier spacing that aggravates the system sensitivity to frequency shift. However, the A-FFT shows a significant advantage in slower increase of MSE. Numerically, it still maintains MSE performance at $-16.51$ dB with $ K=1024$, while the P-FFT and F-FFT cannot work.

\section{Conclusions} \label{Section:Conclusions}
In this paper, a frequency-domain ICI mitigation method named A-FFT is proposed, wherein each recieved symbol is shifted by fractional multiples of the fiducial frequency offset adapted to the time-varying Doppler spread. We have modeled the selection of fiducial frequency offset, and then designed an adaptive estimation algorithm employing the coordinate descent approach. Adopting the idea of the existing F-FFT and combining the designed estimation algorithm, the proposed A-FFT is likely to alleviate ICI effectively for the UWA channel with severe Doppler fluctuations. Compared to the existing methods, i.e., the P-FFT and the F-FFT, simulation results have demonstrated the significant improvements that can be obtained using the proposed technique with medium and high Doppler factors and large carrier numbers.
\bibliographystyle{IEEEtran}
\bibliography{IEEEabrv,LatexWritingModel_Reference}

% Generated by IEEEtran.bst, version: 1.13 (2008/09/30)
\begin{thebibliography}{10}
\providecommand{\url}[1]{#1}
\csname url@samestyle\endcsname
\providecommand{\newblock}{\relax}
\providecommand{\bibinfo}[2]{#2}
\providecommand{\BIBentrySTDinterwordspacing}{\spaceskip=0pt\relax}
\providecommand{\BIBentryALTinterwordstretchfactor}{4}
\providecommand{\BIBentryALTinterwordspacing}{\spaceskip=\fontdimen2\font plus
\BIBentryALTinterwordstretchfactor\fontdimen3\font minus
  \fontdimen4\font\relax}
\providecommand{\BIBforeignlanguage}[2]{{%
\expandafter\ifx\csname l@#1\endcsname\relax
\typeout{** WARNING: IEEEtran.bst: No hyphenation pattern has been}%
\typeout{** loaded for the language `#1'. Using the pattern for}%
\typeout{** the default language instead.}%
\else
\language=\csname l@#1\endcsname
\fi
#2}}
\providecommand{\BIBdecl}{\relax}
\BIBdecl

\bibitem{ChannelEstimateThree}
Y.~Li, S.~Wang, C.~Jin, Y.~Zhang, and T.~Jiang, ``A survey of underwater
  magnetic induction communications: Fundamental issues, recent advances, and
  challenges,'' \emph{IEEE Commun. Surveys Tuts.}, vol.~21, no.~3, pp.
  2466--2487, 2019.

\bibitem{ChannelEstimateFour}
Y.~Li, Y.~Zhang, W.~Li, and T.~Jiang, ``Marine wireless big data: Efficient
  transmission, related applications, and challenges,'' \emph{IEEE Wireless
  Commun.}, vol.~25, no.~1, pp. 19--25, 2018.

\bibitem{channelEstimation}
Z.~{Tang}, R.~C. {Cannizzaro}, G.~{Leus}, and P.~{Banelli}, ``Pilot-assisted
  time-varying channel estimation for {OFDM} systems,'' \emph{IEEE Trans.
  Signal Process.}, vol.~55, no.~5, pp. 2226--2238, 2007.

\bibitem{DopplerRateRange}
X.~Zhuo, M.~Liu, Y.~Wei, G.~Yu, F.~Qu, and R.~Sun, ``{AUV-Aided}
  energy-efficient data collection in underwater acoustic sensor networks,''
  \emph{IEEE Internet Things J.}, vol.~7, no.~10, pp. 10\,010--10\,022, 2020.

\bibitem{fullProposed}
Y.~M. {Aval} and M.~{Stojanovic}, ``Differentially coherent multichannel
  detection of acoustic {OFDM} signals,'' \emph{IEEE J. Ocean. Eng.}, vol.~40,
  no.~2, pp. 251--268, 2015.

\bibitem{PFFTOne}
M.~{Stojanovic}, ``A method for differentially coherent detection of {OFDM}
  signals on {Doppler}-distorted channels,'' in \emph{Proc. IEEE SAM}, 2010,
  pp. 85--88.

\bibitem{PFFTTwo}
S.~{Yerramalli}, M.~{Stojanovic}, and U.~{Mitra}, ``Partial {FFT} demodulation:
  A detection method for doppler distorted {OFDM} systems,'' in \emph{Proc.
  Int. Workshop Signal Process. Adv. Wireless Commun.}, 2010, pp. 1--5.

\bibitem{PFFTThree}
S.~Yerramalli, M.~Stojanovic, and U.~Mitra, ``Partial {FFT} demodulation: A
  detection method for highly doppler distorted {OFDM} systems,'' \emph{IEEE
  Trans. Signal Process.}, vol.~60, no.~11, pp. 5906--5918, 2012.

\bibitem{PFFTFour}
J.~{Yin}, W.~{Ge}, X.~{Han}, B.~{Liu}, and L.~{Guo}, ``Partial {FFT}
  demodulation with {IRC} in {MIMO-SC-FDE} communication over doppler distorted
  underwater acoustic channels,'' \emph{IEEE Comm. Lett.}, vol.~23, no.~11, pp.
  2086--2090, 2019.

\bibitem{frequencyResample}
Z.~{Wang}, S.~{Zhou}, G.~B. {Giannakis}, C.~R. {Berger}, and J.~{Huang},
  ``Frequency-domain oversampling for zero-padded {OFDM} in underwater acoustic
  communications,'' \emph{IEEE J. Ocean. Eng.}, vol.~37, no.~1, pp. 14--24,
  2012.

\bibitem{firstProposed}
Y.~M. {Aval} and M.~{Stojanovic}, ``Fractional {FFT} demodulation for
  differentially coherent detection of acoustic {OFDM} signals,'' in
  \emph{Proc. 46th Asilomar Conf. Signal Syst. Comput.}, 2012, pp. 1525--1529.

\bibitem{DFFFT}
X.~Ma and C.~Zheng, ``Decision fractional fast fourier transform {Doppler}
  compensation in underwater acoustic orthogonal frequency division
  multiplexing,'' \emph{J. Acoust. Soc. Am.}, vol. 140, no.~5, pp.
  EL429--EL433, 2016.

\bibitem{channelModel}
P.~{Qarabaqi} and M.~{Stojanovic}, ``Statistical characterization and
  computationally efficient modeling of a class of underwater acoustic
  communication channels,'' \emph{IEEE J. Ocean. Eng.}, vol.~38, no.~4, pp.
  701--717, 2013.

\end{thebibliography}

\end{document}